\pdfoutput=1
\documentclass[draft]{agujournal}
\draftfalse

\usepackage{epstopdf}
\usepackage{url}

\journalname{JGR-Earth Surface}

\begin{document}
\title{Comment on ``Distinct Thresholds for the Initiation and Cessation of Aeolian Saltation From Field Measurements'' by Raleigh L. Martin and Jasper F. Kok: Alternative Interpretation of Measured Thresholds as two Distinct Cessation Thresholds}
\authors{Thomas P\"ahtz\affil{1,2}}

\affiliation{1}{Institute of Port, Coastal and Offshore Engineering, Ocean College, Zhejiang University, 866 Yu Hang Tang Road, 310058 Hangzhou, China}
\affiliation{2}{State Key Laboratory of Satellite Ocean Environment Dynamics, Second Institute of Oceanography, 36 North Baochu Road, 310012 Hangzhou, China}

\correspondingauthor{Thomas P\"ahtz}{0012136@zju.edu.cn}

\begin{keypoints}
\item Martin and Kok (2018a, \url{https://doi.org/10.1029/2017JF004416}) measure two distinct aeolian transport thresholds: the thresholds of intermittent saltation and continuous saltation
\item Thresholds may be reinterpreted as rebound threshold (intermittent saltation) and impact entrainment threshold (continuous saltation)
\item Preliminary laboratory and observational evidence provides support for alternative interpretation 
\end{keypoints}

\begin{abstract}
Martin and Kok (2018a, \url{https://doi.org/10.1029/2017JF004416}) measured two distinct aeolian saltation transport thresholds: a larger threshold below which continuous saltation transport becomes intermittent and a smaller threshold below which intermittent saltation transport ceases. In the spirit of Bagnold, they interpreted the former threshold as the \textit{fluid threshold}, associated with transport initiation, and the latter threshold as the \textit{impact threshold}, associated with transport cessation. Here I describe and support an alternative interpretation of these two thresholds as two distinct cessation thresholds associated with splash entrainment and, respectively, with compensating energy losses of rebounding particles. This interpretation was recently proposed by P\"ahtz and Dur\'an (2018a, \url{https://doi.org/10.1029/2017JF004580}). To resolve this controversy, further field studies are needed.
\end{abstract}

\section{Introduction}
It is notoriously difficult to extract reliable aeolian saltation transport data from field experiments as the experimental conditions cannot be fully controlled and are disturbed by the measurement instruments. Despite such difficulties, Raleigh L. Martin and Jasper F. Kok managed to provide a rich variety of well-behaving field data with a relatively small scatter in a series of recent studies \citep{MartinKok17,MartinKok18a,MartinKok18b,Martinetal18} thanks to innovative experimental designs. Here I comment on the experiments by \citet{MartinKok18a}, which provided measures for two distinct thresholds values of the wind shear velocity $u_\ast$: a larger threshold (henceforth denoted as $u^e_t$) below which continuous saltation transport becomes intermittent and a smaller threshold (henceforth denoted as $u^r_t$) below which intermittent saltation transport ceases. \citet{MartinKok18a} measured that the ratio $u^r_t/u^e_t$ ranges from about $0.795$ to $0.89$, which is one of the reasons that led them to the interpretation that $u^r_t$ and $u^e_t$ are equal to the cessation threshold $u_{\mathrm{ct}}$ and initiation threshold $u_{\mathrm{it}}$, respectively, of saltation transport. In fact, direct wind tunnel measurements of $u_{\mathrm{ct}}$ and $u_{\mathrm{it}}$ indicate that the ratio $u_{\mathrm{ct}}/u_{\mathrm{it}}$ is in the very same range \citep{Bagnold41,Chepil45,Carneiroetal15}.

In my comment, I describe (section~\ref{Reinterpretation}) and support (section~\ref{SupportingArguments}) an alternative interpretation, recently proposed by \citet{PahtzDuran18a}, of the physical meanings of $u^r_t$ and $u^e_t$.

\section{An Alternative Interpretation of the Measured Transport Thresholds} \label{Reinterpretation}
\citet{PahtzDuran18a} proposed that the thresholds of intermittent ($u^r_t$) and continuous ($u^e_t$) saltation transport are two cessation thresholds associated with two distinct physical interpretations of how saltation transport is sustained by the impacts of particles onto the surface of the sand bed. The first interpretation originates from \citet[][p.~94]{Bagnold41}: ``Physically [the cessation threshold] marks the critical stage at which the energy supplied to the saltating grains by the wind begins to balance the energy losses due to friction when the grains strike the ground [and rebound].'' \citet{PahtzDuran18a} identified the threshold associated with this mechanism as $u^r_t$, like \citet{Bagnold41}, and termed it \textit{rebound threshold}. Bagnold's interpretation of $u^r_t$ was relatively early on replaced by the splash entrainment interpretation \citep[e.g.,][]{Chepil45,Owen64}, which is still mainstream in the aeolian transport community \citep{Duranetal11,Koketal12,Valanceetal15}. This interpretation states that the cessation threshold marks the critical stage at which erosion caused by the splash due to particles impacting onto the bed surface balances deposition. However, \citet{PahtzDuran18a} identified the threshold associated with this mechanism as $u^e_t$ rather than $u^r_t$ and termed $u^e_t$ \textit{impact entrainment threshold}. In fact, these two cessation threshold interpretations, and thus $u^r_t$ and $u^e_t$, are distinct from each other because the rebound mechanism is independent of whether the sand bed is rigid or erodible \citep{Berzietal16}.

The measurements by \citet{MartinKok17,MartinKok18a} indicate that $u^r_t$ controls the rate $Q$ of saltation transport through the law $Q\sim u_\ast^2-u_t^{r2}$ provided that transport is nearly continuous (i.e., $u_\ast\gtrsim u^e_t$), exactly as \citet{PahtzDuran18a} predicted. To demonstrate the conceptual picture by \citet{PahtzDuran18a}, consider the following thought experiment. Assume that saltating particles are subjected to the mean turbulent flow (i.e., without turbulent fluctuations), move in identical periodic trajectories, always rebound in the same manner with the sand bed, and are {\em{never trapped by the bed}}. For this hypothetical case, there is a minimal value of the wind shear velocity $u_\ast$ --- the rebound threshold $u^r_t$ --- below which such a trajectory does not exist despite the absence of deposition \citep{Berzietal16}. That is, for $u_\ast<u^r_t$, saltating particles lose more energy during their rebounds with the bed than the wind supplies during their hops and saltation transport will eventually stop. In contrast, for $u_\ast\geq u^r_t$, saltation transport is continuous and never ceases. However, this is not true for realistic saltation transport because saltating particles are occasionally trapped by the bed, which must be compensated by the entrainment of bed particles through the action of fluid forces and/or particle-bed impacts. It is crucial to emphasize here that the mere occurrence of entrainment is not a sufficiently restrictive requirement because, when a particle acquires an energy below a certain critical value upon entrainment, its initial trajectory is too short so that it gains less energy from the wind than it dissipates when rebounding with the bed, which means that it will rapidly settle as its subsequent trajectories become shorter and shorter. From direct sediment transport simulations, \citet{PahtzDuran18a} found that the wind shear velocity $u^e_t$ at which impact entrainment generates particles with an above-critical energy at a sufficiently high rate is always larger than $u^r_t$ and linked this finding to insufficiently energetic particle-bed impacts for $u^r_t\leq u_\ast<u^e_t$. That is, saltation transport below $u^e_t$ partially requires fluid entrainment to be sustained and is thus intermittent because fluid entrainment only occurs during occasional strong turbulent events \citep{Diplasetal08,Valyrakisetal10,Valyrakisetal13,Pahtzetal18}. Consistently, in numerical simulations that neglect turbulent fluctuations, saltation transport truly stops below $u^e_t$ after a finite period of time (as the mean turbulent flow is much too weak to directly entrain bed particles), which means that $Q$ jumps from a finite value at $u^e_t$ to zero below $u^e_t$ \citep{Carneiroetal11,PahtzDuran18a}.

I would like to emphasize that, according to \citet{PahtzDuran18a}, this alternative interpretation of $u^r_t$ and $u^e_t$ applies to nonsuspended sediment transport driven by an arbitrary Newtonian fluid, which includes viscous and turbulent bedload transport in fluvial environments. In particular, $u^r_t$ controls the transport rate $Q$ of bedload transport through similar laws with the same restriction (i.e., continuous transport and thus $u_\ast\gtrsim u^e_t$) \citep{PahtzDuran18b} and $Q$ also undergoes a discontinuous transition at $u^e_t$ in numerical simulations that neglect turbulent fluctuations around the mean turbulent flow \citep{Clarketal15,Clarketal17,PahtzDuran18a}. The reason for the latter is that impact entrainment dominates mean flow entrainment for all transport regimes but viscous bedload transport \citep{PahtzDuran17} (in viscous bedload transport, $u^e_t\simeq u^r_t$ \citep{PahtzDuran18a}).

\section{Preliminary Evidence Supporting the Interpretation by \citet{PahtzDuran18a}} \label{SupportingArguments}
\subsection{Intermittent Saltation Transport at Initiation Threshold} \label{Windtunnel}
In wind tunnel investigations of beginning and ceasing saltation transport, wind is blown over a flat sand bed at the tunnel surface and the wind speed successively incremented, while sand is either fed or not fed at the tunnel entrance. The threshold $u_{\mathrm{ct}}$ ($u_{\mathrm{it}}$) is then defined as the value of the shear velocity $u_\ast$ at which a cloud of saltating particles can be detected by optical means at the tunnel exit for the feeding (nonfeeding) case. In the nonfeeding case, saltation is initiated due to the action of fluid forces on bed particles (which is why $u_{\mathrm{it}}$ has been called \textit{fluid threshold} \citep{Bagnold41}), while saltation already occurs in the feeding case and is sustained at lower wind speed due to particles impacting onto the bed surface (which is why $u_{\mathrm{ct}}$ has been called \textit{impact threshold} \citep{Bagnold41}).

The only study in which $u_{\mathrm{it}}$ and $u_{\mathrm{ct}}$ and saltation transport intermittency characteristics have been simultaneously measured is the one by \citet{Carneiroetal15}. These authors reported that $u_{\mathrm{it}}/u_{\mathrm{ct}}\approx0.85$ and that saltation transport at both thresholds is intermittent (i.e., the fraction of time that saltation transport occurs is significantly smaller than unity). The latter finding contradicts the interpretation by \citet{MartinKok18a} that the threshold $u^e_t$ of continuous transport is equal to the initiation threshold $u_{\mathrm{it}}$. However, one has to keep in mind that the manner in which saltation transport was recorded by \citet[][optical measurements]{Carneiroetal15} differed from the one by \citet[][sand trap measurements]{MartinKok18a}, which is crucial because the distinction between intermittent and continuous transport strongly depends on the spatial and temporal scales of sampling and the sensitivity of the saltation transport measurements. Likewise, \citet{Carneiroetal15} also measured the threshold of continuous saltation transport, which turned out to be not too far from the initiation threshold: $u^e_t\approx1.1u_{\mathrm{it}}$. I therefore do not think that this evidence is sufficiently strong to rule out $u^e_t=u_{\mathrm{it}}$, but it causes some doubt.

\subsection{Different Saltation Transport Initiation in Laboratory and Field} \label{TunnelField}
In wind tunnel experiments, the initiation of saltation transport follows a common pattern. When the wind shear velocity $u_\ast$ exceeds the initiation threshold $u_{\mathrm{it}}$, saltation transport occurs at the downwind end of the test section and is preceded by a rolling regime (which includes sporadic small particle hops) further upwind \citep{Bagnold41,Burretal15}. Based on theoretical arguments and wind tunnel measurements, \citet{Pahtzetal18} put forward that rolling transport is the more readily initiated the larger the thickness $\delta$ of the turbulent boundary layer because $\delta$ controls the size of the largest turbulent eddies \citep{AlhamdiBailey17}. For boundary layers produced in typical wind tunnel experiments (comparably small $\delta$), the rolling threshold is larger than for atmospheric boundary layers in the field (comparably large $\delta$). In particular, in wind tunnels, the rolling threshold is significantly larger than the threshold at which saltation transport can be sustained ($u_{\mathrm{ct}}$), which is why rolling transport evolves into saltation transport. In contrast, in the field, the rolling threshold may be smaller than $u_{\mathrm{ct}}$ and rolling transport may therefore not evolve into saltation transport. In fact, there are observational hints that some kind of aeolian sand transport may, indeed, occur below $u_{\mathrm{ct}}$ in the field, such as wind erosion of gravel on Earth \citep{DeSilvaetal13} and a larger-than-expected mobility of coarse particles on Mars \citep{Bakeretal18}. If the case for sub-$u_{\mathrm{ct}}$ transport in the field was to become stronger in the future, it would likely imply that initiation and cessation of saltation transport in the field are equivalent ($u_{\mathrm{ct}}=u_{\mathrm{it}}$) rather than distinct ($u_{\mathrm{ct}}<u_{\mathrm{it}}$). In that case, the thresholds $u^r_t$ and $u^e_t$ measured by \citet{MartinKok18a}, as they are distinct from each other ($u^r_t<u^e_t$), cannot be the same as $u_{\mathrm{ct}}$ and $u_{\mathrm{it}}$, respectively.
 
\section{Conclusions}
\citet{MartinKok18a} interpreted their measured thresholds of intermittent ($u^r_t$) and continuous ($u^e_t$) saltation transport as impact threshold and fluid threshold, respectively. I have described the alternative interpretation by \citet{PahtzDuran18a} that $u^r_t$ and $u^e_t$ are the rebound threshold and impact entrainment threshold, respectively. Currently, the evidence is not sufficiently conclusive to falsify any of these two interpretations. Therefore, further field studies, such as direct measurements of the aerodynamic entrainment of sand bed particles in the field, are needed to resolve this controversy.

\acknowledgments
The experimental data to which this comment refers are provided by \citet{MartinKok17,MartinKok18a}. I acknowledge support from grant National Natural Science Foundation of China (No.~11750410687).


\end{document}